\documentclass{aa}
\usepackage{graphicx}
\usepackage{times}
\usepackage{astron}
\usepackage{psfig}

\def\hide#1{}

\def\eg{{e.g.}\/}

\newcommand{\apr}{$\approx$}

\begin{document}

\thesaurus{03.13.2; 03.20.4; 04.19.1; 11.17.3; 12.05.1}

\title{Did Most High-Redshift Quasars Escape Detection?}

\author{C. Wolf\inst{1} \and K. Meisenheimer\inst{1} \and H.-J. R\"oser\inst{1} 
\and S.V.W. Beckwith\inst{1,3} \and R. Fockenbrock\inst{1} \and H. Hippelein\inst{1} \and B. von Kuhlmann\inst{1} \and S. Phleps\inst{1} \and E. Thommes\inst{2} }


\institute{
 Max-Planck-Institut f\"ur Astronomie, K\"onigstuhl 17,
   D-69117 Heidelberg, Germany
\and
 Royal Observatory, Blackfort Hill, Edinburgh EH9 3HJ, UK
\and
 Space Telescope Science Institute, 3700 San Martin Drive, Baltimore, MD 21218, USA}

\date{Received 4 Nov 1998 / Accepted 17 Dec 1998}
\maketitle


\begin{abstract}
We present follow-up spectroscopy to the Calar Alto Deep Imaging Survey (CADIS), reporting on the quasars in the CADIS 16\,h-field, and including preliminary data from two other fields. In this $10\arcmin \times 10\arcmin$ field we found six quasars at redshift $z>2.2$ which are brighter than $R=22^m$, while surface densities determined in previous surveys are reported to be below one on average. This is the highest reported surface density of high-redshift quasars within the given limits. We believe this result to be a consequence of the multicolor database and the search technique used in CADIS.

Our observations further seem to indicate that the density of high-redshift quasars is also higher than predicted by the luminosity function of Warren (1994) by a factor of three, implying that common search methods have overlooked the majority of high-redshift quasars at this magnitude level. In particular, an application of the $V/V_{max}$ test to our preliminary list of AGNs with $R<22^m$ and $z<4$ yields $\langle V/V_{max} \rangle = 0.51\pm0.06$, characterizing our preliminary redshift distribution as flat.
\end{abstract}

\section{Introduction}

Searches for high-redshift quasars have been undertaken by many groups using a variety of methods, such as color criteria, the search for emission lines in objective prism spectra or variability \cite{Warren,DMS2,DMS3,SSG,HV}. Almost independent of the method, observed surface densities resulting from deep surveys down to $B \approx 22$ roughly agree \cite{HS}.

Quasars at redshifts above $z=2.2$ have been observed with typically one quarter of the surface density at lower redshifts (see review of Hartwick \& Shade 1990; Hazard 1990; assorted contributions within Crampton 1991; Shaver et al.\ 1996). Since any conclusion about the development of the luminosity function of quasars depends on the completeness of the surveys, there has been a lot of discussion about the completeness of search methods for high-redshift quasars, which differ from the methods used at low redshift.

In this paper we present first results for high-redshift quasars in the Calar Alto Deep Imaging Survey (CADIS), which is outlined in Sect. 2. Section 3 presents a brief description of our quasar search method and the spectroscopic confirmation of our candidates. As discussed in Sect. 4 and 5, CADIS finds high-redshift quasars with unprecedented completeness and above the density revealed by previous work. 

\section{The Calar Alto Deep Imaging Survey}

\begin{figure*}
\centerline{\psfig{figure=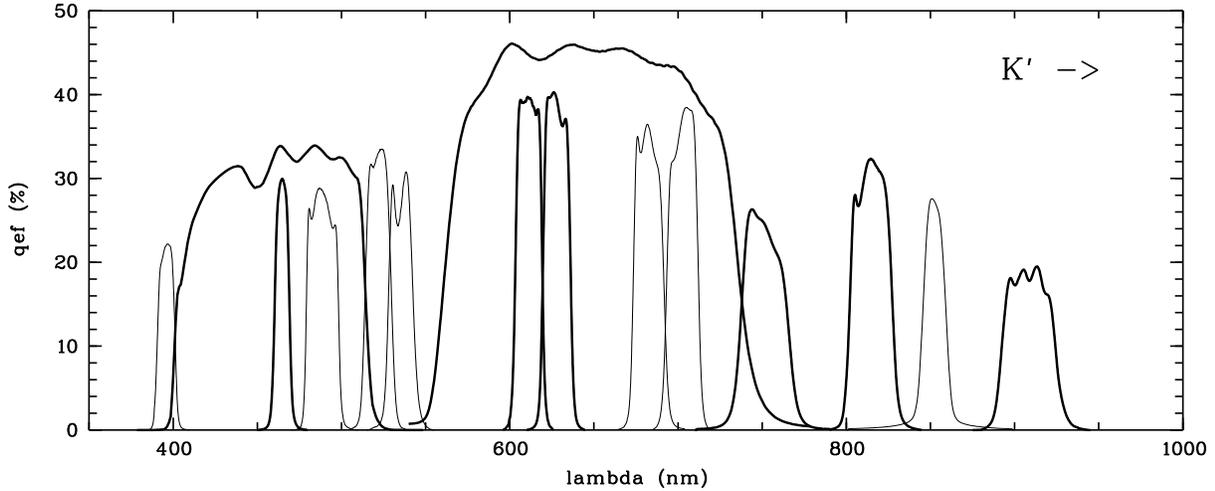,angle=270,clip=t,width=16cm}}
\caption[ ]{The full CADIS filterset contains three non-standard broadband filters, $B$, $R$ and $K^\prime$ (not shown), and 13 mediumband filters providing a lot of useful data for multicolor classification. Filters already observed on the 16\,h-field are highligthed by thick lines. \label{filterset}}
\end{figure*}

The most prominent motivation for CADIS is a new concept for finding primeval galaxies \cite{meise}. CADIS combines an emission line survey carried out with a Fabry-P\'erot interferometer and a deep multicolor survey using three broadband filters ($B$, $R$ and $K^\prime$) and thirteen medium-band filters in the wavelength range between 390 nm and 930 nm when fully completed (see Fig.\,\ref{filterset}). This combination allows to adress many pending questions concerning a broad variety of objects based on the same dataset. We obtain a lot of multicolor data about faint galactic stars \cite{P,CW1} as well as quasars and galaxies over a broad range of redshift. CADIS is supposed to produce candidates for quasars and Seyfert galaxies to magnitudes as faint as $R=23^m$. Also, the accompanied deep $K^\prime$-band survey establishes one of the largest deep $K^\prime$-band galaxy surveys \cite{JH} and allows to search for unusually red objects \cite{djt}. 

The nine CADIS fields measure \apr 100\,$\sq\arcmin$ each and are all located at high Galactic latitude ($b\ge45\degr$) in order to avoid extinction and reddening of object colors. We included the 3\,h-field of the Canadian French Redshift Survey \cite{cfrs_3h} as a ``tenth'' CADIS field to have spectroscopic results for one field readily available. Observations are most complete in the 16\,h-field, where CCD photometry in nine bands is almost finished (see Table\,\ref{filtertab}). Observations are performed with the focal reducers CAFOS (Calar Alto Faint Object Spectrograph) at the 2.2-m-telescope and MOSCA (Multi-Object Spectrograph for Calar Alto) and OMEGAprime \cite{bizenberger} at the 3.5-m-telescope at Calar Alto Observatory, Spain.

For photometric calibration, we establish a system of faint standard stars in the CADIS fields, which are calibrated with respect to spectrophotometric standards in photometric nights \cite{oke}. This way we have several spectrophotometric standards within each CADIS exposure providing us with independence from photometric conditions for regular imaging. CCD images are reduced such that accurate color measurements are ensured even under conditions of changing seeing. The reduction procedure is outlined in R\"oser \& Meisenheimer (1991). Between all wavebands, the relative calibration is better than 3\% for objects of $R=22^m$.

\begin{table}
\caption{Filters observed on the CADIS 16\,h-field and used for object classification. Exposure times are all for the 2.2-m-telescope. For observations done at the 3.5-m-telescope, we list the equivalent exposure at the 2.2-m-telescope. The $10\sigma$ limiting magnitude is estimated from the magnitude distribution of those objects with flux errors measured to be roughly 10\%. Some filters need a little more exposure to reach the depth we aim for, $m_{lim,goal}$. \label{filtertab}}
\begin{tabular}{lrrr}
$\lambda_{cen}$/fwhm (nm) & $t_{exp}$/sec & $m_{lim,10\sigma}$ & $m_{lim,goal}$ \\
\noalign{\smallskip}
\hline
\noalign{\smallskip}
465/10  		& 20000 & 24.1 & 23.8 \\
461/113 ($B$)		&  6200 & 24.7 & 25.1 \\
611/16  		& 11000 & 23.4 & 23.4 \\
628/16  		& 23000 & 23.6 & 23.4 \\
649/172 ($R$)		&  5300 & 24.1 & 23.7 \\
752/28  		& 15600 & 22.9 & 23.0 \\
815/25  ($I$)		& 30700 & 22.9 & 22.9 \\
909/31  		& 31600 & 22.3 & 22.5 \\
2120/340 ($K^\prime$)	&  9000 & 19.5 & 19.7 \\
\noalign{\smallskip}
\hline
\end{tabular}
\end{table}

\section{Finding and probing quasar candidates}

\subsection{Multicolor classification and redshift estimation}

For CADIS we opted for a classification scheme, that essentially compares observed colors of each object with a color library of known objects assembled from observed spectra by synthetic photometry performed on our CADIS filter set. The spectral libraries used as an input were the Gunn \& Stryker (1983) catalogue for stars, the galaxy template spectra from Kinney et al. (1996), and the QSO template spectrum of Francis et al. (1991). From this, we generated regular grids of QSO templates ranging in redshift within $0<z<4$ and having various continuum slopes and emission line equivalent widths. Also, a grid of galaxy templates has been generated for $0<z<2$, and contains various spectral types from old populations to starbursts. We have not attempted to find quasars at $z>4$, yet, since they are believed to be extremely rare and we were not yet able to model their colors well enough. 

Objects are classified by locating them in color space and comparing the probability for each class to generate the given measurement. Given the photometric error ellipsoid in the n-dimensional color space, each library object can be assigned a probability to cause an observation of the measured colors. For a whole class, this probability is assumed to be the average value of the individual class members (Parzen's Kernel estimator, 1963). From these probabilities we can derive the likelihood of each object to belong to the various classes.

Since the galaxy and quasar libraries resemble regular grids in redshift and spectral type, these parameters can also be estimated from the observation. For this purpose, we treat the library as a statistical ensemble generating the measured colors and determine expectation values as well as variances for assessing the quality of the estimate. A detailed discussion of the algorithm and the performance of the classification and redshift estimation will be given in Wolf (1998).

Up to now, we select quasar candidates on the basis of both morphological and color criteria, because with the presently available part of the filterset there is some overlap of low-redshift galaxies with high-redshift quasars left in our color space, which scatters a tiny fraction of the true galaxies into the quasar class. In order to sort out these resolved contaminating galaxies, only point sources with colors resembling most likely those of quasars are considered candidates. We might miss resolved low luminosity quasars using this morphological criterion, but simulations tell us that the class ambiguity should be removed by completion of the entire filterset. Then we should be able to drop the morphological criterion.

Colors resembling {\it most likely} those of quasars means in our case that an object has been assigned a likelihood of more than 75\% to be a member of the quasar class. For each object we normalized the likelihoods such, that the three object classes would always add up to 100\%. In the 16\,h-field we found eight quasar candidates brighter than $R=22$ with $z<4$. Although, the 9\,h-field and the 3\,h-field data do not have the same depth and quality yet, they show a similar number of candidates.

\begin{figure}
\centerline{\psfig{figure=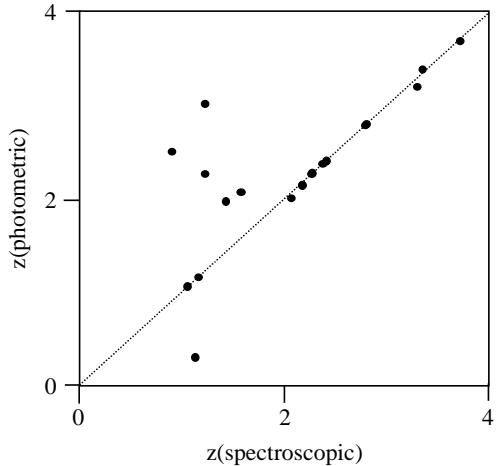,clip=t,width=6.5cm}}
\caption[ ]{The use of many mediumband filters enables us to estimate the redshift of quasars. The performance is best at $z>2$. \label{qso_zz}}
\end{figure}

\subsection{Spectroscopic verification}

All quasar candidates on the 16\,h-field except one were observed among other objects with the multi-object spectrograph MOSCA at the 3.5-m-telescope on Calar Alto. These other objects were galactic stars and galaxies at various redshifts including candidates for Seyfert galaxies, which were observed in order to check the quality of our classification algorithm. One candidate was observed with longslit spectroscopy at the 2.2-m-telescope on Calar Alto, because its position within the field did not allow inclusion onto the multi-object masks.

Between July 4, 1997 and July 8, 1997 we exposed three multi-slit masks for 4000\,s, 8000\,s and 12000\,s, respectively. Most spectra cover the wavelength range of 4000--9500\,\AA\ at 12\,\AA\ resolution given by the slit width of $1\farcs75$, and some cover the range of 4800--10000\,\AA. Generally, the grism G\,500 blazed at 5640\,\AA\ was used, providing a wavelength scale of 2.8\,\AA/pixel on a $2k$ x $2k$-Loral CCD with $15\mu$ pixel size. Quasar 16\,h-1610 was left out from the masks and observed on July 3, 1997 using grism green-200 of CAFOS at the 2.2-m-telescope with a wavelength coverage of 4000--8500\,\AA. The exposure of 1500\,s allowed the identification of three broad emission lines despite a low S/N ratio for the continuum.

\begin{figure}
\centerline{\psfig{figure=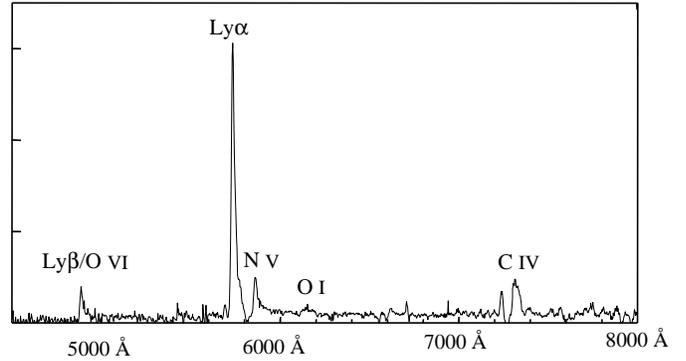,clip=t,width=8.8cm}}
\caption[ ]{The broad absorption line quasar CADIS 16\,h-780 at redshift 3.72 shows an unusually strong Lyman $\alpha$-line. This spectrum is not flux-calibrated and plotted in arbitrary count units. \label{spec_780}}
\end{figure}

\begin{table*}
\caption{Quasars found in the CADIS fields and confirmed by spectroscopy. The blue luminosity $M_{B}$ has been calculated for $H_0 = 50$ km sec$^{-1}$ Mpc$^{-1}$ and $q_0 = 0.5$. For low-$z$ objects, the luminosity is calculated from the far-red photometry, while for $z>1.1$-objects a spectral index is derived from the far-red and $K^\prime$-photometry (not available on the 3\,h-field). This way, luminosities can be estimated quite well even for $z \approx 4$-objects. The last two quasars from the northern part of the 3\,h-field were identified by the CFRS. \label{quasars}}
\begin{flushleft}
\begin{tabular}{llllllll}
\hline
\noalign{\smallskip}
$\alpha_{2000}$ & $\delta_{2000}$ & $R$ & $z_{\rm multicolor}$ & $z_{\rm spec}$ & $M_{B}$ & Nr & Spectral features identified \\
\noalign{\smallskip}
\hline
\noalign{\smallskip}
$16^{h}24^{m}26\fs31$ & $55\degr 47\arcmin 49\farcs7$ & 20\fm11 & n.a.		  & 0.47 & $-21.7$ & 644 & $H\beta$ 4861, [O\,{\sc iii}] 4959, [O\,{\sc iii}] 5007, $H\alpha$ 6563 \\
$16^{h}24^{m}09\fs41$ & $55\degr 44\arcmin 47\farcs2$ & 20\fm89 & $0.32 \pm 0.01$ & 1.13 & $-22.6$ & 2028 & C\,{\sc iii}] 1909, Mg\,{\sc ii} 2799 \\
$16^{h}24^{m}47\fs36$ & $55\degr 48\arcmin 38\farcs1$ & 22\fm35 & $1.99 \pm 0.10$ & 1.43 & $-22.2$ & 604 & C\,{\sc iii}] 1909, Mg\,{\sc ii} 2799 \\
$16^{h}24^{m}24\fs15$ & $55\degr 42\arcmin 26\farcs8$ & 21\fm58 & $2.08 \pm 0.01$ & 1.57 & $-23.4$ & 429 & C\,{\sc iii}] 1909, C\,{\sc ii}] 2326, Ne\,{\sc iv} 2423, Mg\,{\sc ii} 2799 \\
$16^{h}23^{m}59\fs10$ & $55\degr 41\arcmin 08\farcs9$ & 21\fm53 & $2.27 \pm 0.02$ & 2.26 & $-23.8$ & 1610 & Si\,{\sc iv}+O\,{\sc iv}] $\sim$1400, C\,{\sc iv} 1549, C\,{\sc iii}] 1909 \\
$16^{h}24^{m}21\fs19$ & $55\degr 42\arcmin 43\farcs3$ & 21\fm84 & $2.29 \pm 0.01$ & 2.27 & $-23.6$ & 1373 & Si\,{\sc iv}+O\,{\sc iv}] $\sim$1400, C\,{\sc iv} 1549, C\,{\sc iii}] 1909 \\
$16^{h}24^{m}59\fs82$ & $55\degr 42\arcmin 28\farcs1$ & 18\fm68 & $2.41 \pm 0.03$ & 2.41 & $-26.6$ & 330 & Si\,{\sc iv}+O\,{\sc iv}] $\sim$1400, C\,{\sc iv} 1549, He\,{\sc ii} 1640, C\,{\sc iii}] 1909 \\
$16^{h}24^{m}12\fs94$ & $55\degr 44\arcmin 46\farcs8$ & 20\fm25 & $2.80 \pm 0.01$ & 2.80 & $-24.9$ & 540 & Si\,{\sc iv}+O\,{\sc iv}] $\sim$1400, C\,{\sc iv} 1549, He\,{\sc ii} 1640, C\,{\sc iii}] 1909 \\
$16^{h}24^{m}13\fs67$ & $55\degr 48\arcmin 34\farcs8$ & 21\fm97 & $3.38 \pm 0.01$ & 3.36 & $-24.1$ & 1382 & Ly\,$\beta$+O\,{\sc vi} $\sim$1031, Ly\,$\alpha$+N\,{\sc v} 1216, C\,{\sc iv} 1549 \\
$16^{h}24^{m}48\fs86$ & $55\degr 40\arcmin 53\farcs3$ & 21\fm15 & $3.69 \pm 0.01$ & 3.72 & $-26.1$ & 780 & Ly\,$\beta$+O\,{\sc vi} $\sim$1031, Ly\,$\alpha$+N\,{\sc v} 1216, C\,{\sc iv} 1549 (BAL)\\
\noalign{\smallskip}		   	     	    
\hline         	    		   	     	    
\noalign{\smallskip}		   	     	    
$09^{h}13^{m}31\fs71$ & $46\degr 07\arcmin 25\farcs9$ & 18\fm89 & n.a.		 & 0.82 & $-23.9$ & 104 & Mg\,{\sc ii} 2799, Ne\,{\sc v} 3426 (outside 100\,$\sq\arcmin$ field) \\
$09^{h}14^{m}13\fs86$ & $46\degr 16\arcmin 12\farcs9$ & 20\fm99 & $1.17 \pm 0.18$ & 1.16 & $-22.2$ & 1345 & Mg\,{\sc ii} 2799 \\
$09^{h}13^{m}21\fs74$ & $46\degr 11\arcmin 11\farcs0$ & 22\fm10 & $2.28 \pm 0.02$ & 1.22 & $-21.9$ & 781 & Mg\,{\sc ii} 2799 \\
$09^{h}13^{m}25\fs18$ & $46\degr 12\arcmin 15\farcs3$ & 21\fm50 & $3.02 \pm 0.01$ & 1.22 & $-22.2$ & 887 & Mg\,{\sc ii} 2799 \\
$09^{h}14^{m}03\fs65$ & $46\degr 15\arcmin 49\farcs9$ & 19\fm94 & $2.16 \pm 0.01$ & 2.18 & $-25.2$ & 1 & Si\,{\sc iv}+O\,{\sc iv}] $\sim$1400, C\,{\sc iv} 1549, C\,{\sc iii}] 1909 \\
$09^{h}14^{m}04\fs14$ & $46\degr 10\arcmin 44\farcs9$ & 21\fm05 & $2.16 \pm 0.01$ & 2.18 & $-24.3$ & 984 & Si\,{\sc iv}+O\,{\sc iv}] $\sim$1400, C\,{\sc iv} 1549, C\,{\sc iii}] 1909 \\
$09^{h}14^{m}10\fs26$ & $46\degr 10\arcmin 49\farcs8$ & 20\fm23 & $2.33 \pm 0.01$ & 2.37 & $-25.1$ & 822 & Si\,{\sc iv}+O\,{\sc iv}] $\sim$1400, C\,{\sc iv} 1549, C\,{\sc iii}] 1909 (BAL) \\
$09^{h}13^{m}39\fs75$ & $46\degr 17\arcmin 00\farcs1$ & 22\fm56 & $2.77 \pm 0.02$ & 2.81 & $-23.2$ & 320 & Ly\,$\alpha$+N\,{\sc v} 1216, C\,{\sc iv} 1549, C\,{\sc iii}] 1909 \\
\noalign{\smallskip}		   	     	    
\hline         	    		   	     	    
\noalign{\smallskip}		   	     	    
$03^{h}02^{m}43\fs55$ & $00\degr 05\arcmin 08\farcs5$ & 20\fm75 & $2.51 \pm 0.01$ & 0.90 & $-22.4$ & 224 & Mg\,{\sc ii} 2799 \\
$03^{h}02^{m}42\fs81$ & $00\degr 07\arcmin 15\farcs8$ & 18\fm32 & $0.02 \pm 0.01$ & 1.03 & $-25.1$ & 238 & Mg\,{\sc ii} 2799, C\,{\sc iii}] 1909, C\,{\sc ii}] 2326, Ne\,{\sc iv} 2423 \\
$03^{h}02^{m}22\fs08$ & $00\degr 06\arcmin 30\farcs9$ & 20\fm07 & $3.20 \pm 0.01$ & 3.30 & $\approx -26$ & 211 & Ly\,$\alpha$+N\,{\sc v} 1216, C\,{\sc iv} 1549 (BAL)\\
\noalign{\smallskip}		   	     	    
\hline         	    		   	     	    
\noalign{\smallskip}		   	     	    
$03^{h}02^{m}33\fs17$ & $00\degr 13\arcmin 31\farcs2$ & 21\fm04 & $1.08 \pm 0.02$ & 1.05 & $-23.0$ & 9 & identified by CFRS \\
$03^{h}02^{m}23\fs07$ & $00\degr 13\arcmin 12\farcs5$ & 20\fm58 & $2.02 \pm 0.01$ & 2.07 & $\approx -25$ & 178 & identified by CFRS \\
\noalign{\smallskip}
\hline
\end{tabular}
\end{flushleft}
\end{table*}

Spectroscopy was carried out for further checks of our classification procedure also on the CADIS 9\,h-field and the 3\,h-field of the Canadian French Redshift Survey (CFRS), which we had imaged only in a subset of our filters. Between February 26, 1998 and March 3, 1998 we exposed three multi-slit masks on the 9\,h-field for 8000\,s, 7500\,s and 7500\,s, respectively, and one mask on the 3\,h-field for 2000\,s at an average airmass of 2.0, using again the G\,500 grism of MOSCA and 1\farcs5 slits for all of these observations. Object 3\,h-238 was observed on February 22, 1998 using grism blue-200 of CAFOS with a wavelength coverage of 3500--7000\,\AA\ and exposed for 1200\,s. 

\begin{figure*}
\centerline{\hbox{
\psfig{figure=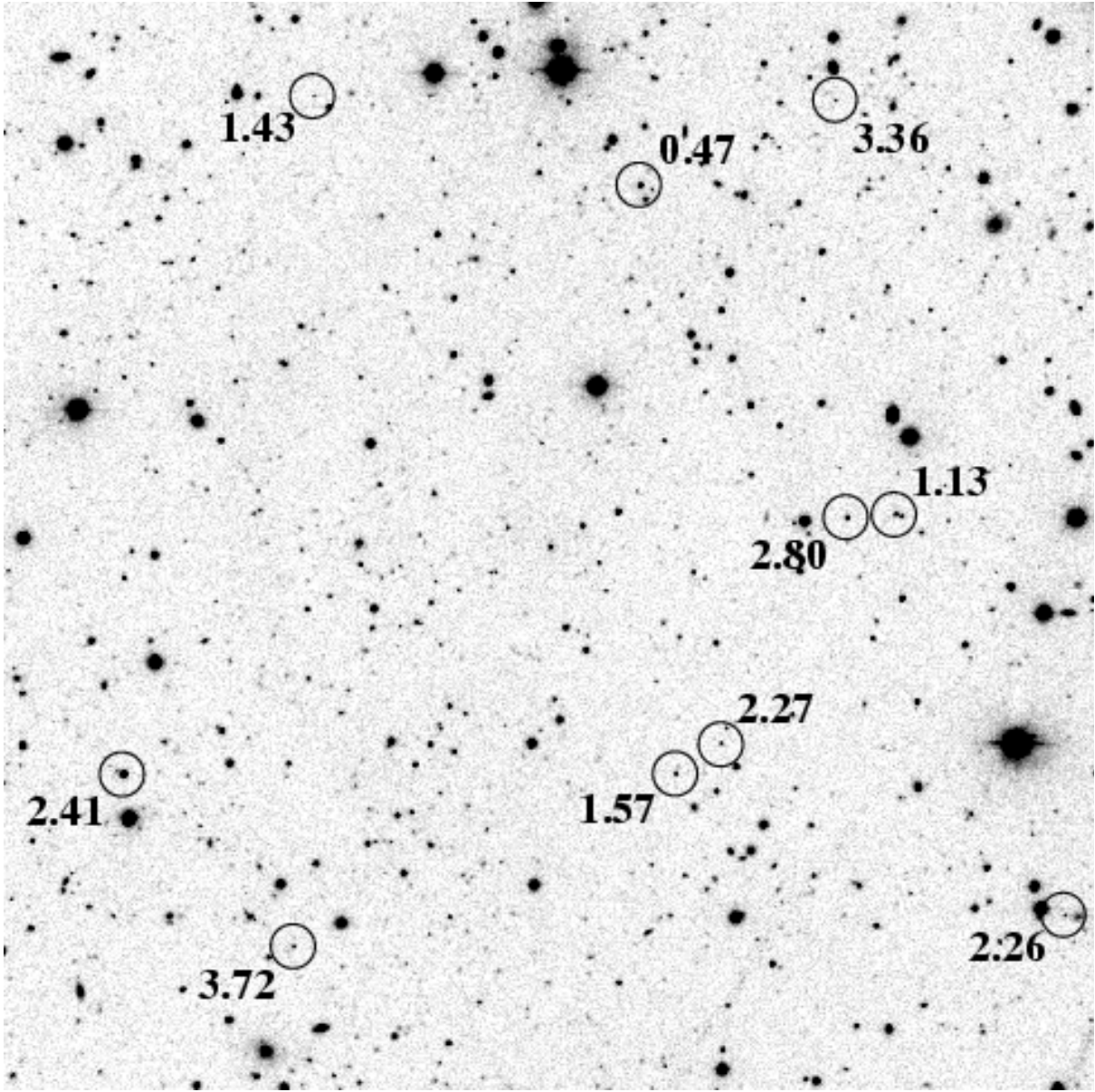,clip=t,width=16cm}}}
\caption[ ]{This $R$ band finding chart shows the quasars and Seyfert\,1 galaxies found in the CADIS 16\,h-field. The field measures $10\arcmin \times 10\arcmin$. North is up and east is to the left. Redshifts are indicated in the image and are mostly above $z=2.2$. \label{16h_field}}
\end{figure*}

On the CADIS 16\,h-field this follow-up spectroscopy confirmed all quasar and Seyfert candidates (see Fig.\,\ref{16h_field}). At $R<22$ and $z<4$, there are no more quasar candidates left to check. Due to incomplete data reduction on the other two fields, we were in spring 1998 only able to check a subset of the candidates we identified by now. Eight candidates were observed and confirmed on the 9\,h-field with four of $R<22$ being left. On the 3\,h-field three candidates were observed and confirmed with five being left to check. Two of the remaining candidates have been observed spectroscopically by the CFRS \cite{cfrs_3h}. One of them turned out to be a quasar at $z=2.07$ and the other one a compact galaxy at $z=0.61$ --- our first misclassified quasar candidate. 

Our redshift estimation yields mostly good results, preferably at true redshifts of $z>2$ (see Fig.\,\ref{qso_zz}). We have performed Monte-Carlo simulations using the model quasars from the library and diffusing their photometry according to the S/N ratios achieved in the CADIS filters. We are able to assess completeness levels and the expected quality of the redshift estimates. The simulations agree very well with the observations, since they show a lot of redshift confusion for $z<2.5$ and accurate estimates at higher redshift. At lower redshifts the quasar spectra do not display features discriminating sufficiently in our filters for a successful estimation. At higher redshift, the continuum break at the Lyman-$\alpha$ line is the strongest spectral feature that can be detected with our filterset. According to the simulations, we should not miss more than 10\% of the library-type quasars with $R<22^m$ \cite{CW2}.

\subsection{The preliminary CADIS quasar sample}

Table\,\ref{quasars} lists all quasars that have been confirmed spectroscopically by CADIS including two quasars identified by the CFRS in its 3\,h-field. Although, the candidate identification on the 9\,h-field and the 3\,h-field is incomplete, we report the present status of our findings. On the CADIS 16\,h-field we found seven quasars brighter than $R=22$ and three Seyfert\,1 galaxies. These ten AGN are more or less uniformly spread over the field. The positions have an accuracy of $\pm$0\farcs1 in each coordinate, measured relative to secondary standards derived from the POSS-II plates and PPM \cite{ppm} stars.

In the 3\,h-field, spectra of 271 objects with $17\fm5<I_{AB}<22\fm5$ were taken as part of the CFRS \cite{cfrs_3h}. This is roughly equivalent to $17^m<I<22^m$ on a magnitude scale, for which Wega defines the zero point. The CFRS selected less than half of the objects present on the field within these limits for spectroscopy with the choice purely based on position. Two quasars at redshifts of 1.05 and 2.07 were found by the CFRS. The CFRS left more than half of the objects unstudied, thereby missing a relatively bright quasar ($R=20\fm1$) at redshift 3.3 and an $18^{th}$ magnitude quasar at $z=1.03$. Among our quasar candidates inside the original CFRS field, we checked only these two so far. Obviously, we took only spectra of objects not studied by the CFRS already. Also, our 3\,h-field is positionally not precisely overlapping with the CFRS field.

Since CADIS is a rather deep survey on a relatively small area compared with other quasar surveys, the objects found by CADIS have intrinsically faint luminosities. All quasars presented are less luminous than $M_B = -27^m$. We note, that none of the quasars presented was detected as a radio source by the VLA FIRST survey, which has covered these three fields to a sensitivity limit of 1\,mJy at 1.47\,GHz \cite{vlafirst}.

\subsection{Notes on individual objects}

Object 16\,h-780 is a broad absorption line (BAL) quasar with a rather strong Lyman $\alpha$ line (see Fig.\,\ref{spec_780}) and the most distant object identified in CADIS, up to now. 

Object 16\,h-540 at redshift 2.80 had in fact already been identified as a good quasar candidate in a variability study, where we compared $R$ magnitudes in frames taken at different epochs throughout the CADIS observing program.

The Seyfert galaxy 16\,h-644 was found by chance among the galaxies we selected for checking our photometric redshift estimates. It has stellar shape and was classified as a starburst galaxy by our algorithm. 

The quasar 16\,h-429 looks morphologically stellar in some filters and extended in others, \eg \ in the $R$ filter and in the $K^\prime$ band. The point source appears to be not precisely centered on the fuzz. Our classification assigned a significant likelihood to this object for being a galaxy at $z\approx1.7$. We can not decide at this point, whether the visible fuzz around this object is a luminous host galaxy or a foreground object projected onto the line of sight by chance. 

Interesting to note are three pairs of AGNs: 
\begin{itemize}
\item the quasar pair 16\,h-1610/1373 at redshift 2.26 and 2.27 with an angular separation of $195\arcsec$, corresponding to a projected lateral distance of $d_p = 1.55$\,Mpc for $H_0 = 50$ km sec$^{-1}$ Mpc$^{-1}$ and $q_0 = 0.5$. 
\item the quasar pair 9\,h-1/984 at redshift 2.18 with an angular separation of $305\arcsec$ ($d_p = 2.45$\,Mpc).
\item the pair of Seyfert\,1 galaxies 9\,h-781/887 at $z=1.22$ being $74\arcsec$ apart ($d_p = 0.64$\,Mpc). 
\end{itemize}
Object 9\,h-104 lies outside of the 100\,$\sq\arcmin$ field, where data is available from all filters. Instead, for this object only four bands taken with CAFOS on its larger field of view were available. In this border area most objects are unclassifiable due to the lack of more data, but object 9\,h-104 was still classified as a quasar.

\section{The observed surface density of quasars}

\begin{table*}
\caption{Cumulative surface densities of quasars as reported in the literature and found in CADIS. The values in the last line are based on the assumption that three of the four remaining candidates in the 9\,h-field were in fact located at $z>2.2$ as suggested by our photometric redshift estimates. The luminosity function by Warren et al. (1994) is used for a predicition under the assumption of 100\% observational completeness. \label{density}}
\begin{tabular}{llllll}
\hline
\noalign{\smallskip}
Authors		   & Year	& Method   & Magnitude   		& Redshift & QSO/$\sq\degr$ \\
\noalign{\smallskip}
\hline
\noalign{\smallskip}
Hartwick \& Schade & 1990	& review      & $B<22\fm5$ 		& $z>2.2$ & $30^{\,+24}_{-12}$ \\
Hawkins \& Veron (extrapolation)  & 1995	& variability & $B<22\fm5$   		& $z>2.2$ & \apr 20 \\
Hall et al. (DMS)  & 1996	& multicolor  & $R<22^m$, $B<22\fm6$	& $z>2.3$ & $15\pm4$ \\
Warren, Hewett, Osmer & 1994	& luminosity function & $R<22^m$   & $z>2.2$ & \apr 68 \\
\noalign{\smallskip}
\hline
\noalign{\smallskip}
CADIS (16\,h-field)        & 1998	& multicolor  & $R<22^m$   & $z>2.2$ & $215\pm90$ \\
CADIS (16\,h+9\,h-field+candidates) & 1998	& multicolor  & $R<22^m$  & $z>2.2$ & $180\pm55$ \\
\noalign{\smallskip}
\hline
\end{tabular}
\end{table*}

Based on the quasars we have spectroscopically confirmed, we calculate a cumulative surface density for quasars ($M_{B}<-23^m$) down to $R=22^m$. In the 16\,h-field we found one object at $z<2.2$ and six objects at $z>2.2$. Since a CADIS field covers 100\,$\sq\arcmin$ on the sky, this translates into a surface density as high as $215\pm90$ QSO/$\sq\degr$ for $R<22^m$ at $z>2.2$ with the error being based on pure Poisson statistics. 

We have not found as many quasars on the other fields, yet, but there we have not completed our observations. If the remaining candidates prove to be quasars at a continuing rate of success, the high density found on the 16\,h-field would not appear unusual. For the fields 9\,h and 16\,h combined, we would get $90\pm40$ QSO/$\sq\degr$ at low redshift ($z<2.2$) and $180\pm55$ QSO/$\sq\degr$ at high redshift ($z>2.2$). In the following, we compare this result with surface densities found by other authors. 

In the literature, a division between low-redshift and high-redshift quasars is set at $z=2.2$, since at this value the Lyman-$\alpha$ line moves from the U band into B band. At $z>2.2$, the U band probes the depressed quasar spectrum bluewards of Lyman-$\alpha$, causing common UV excess searches to fail.

The review article of Hartwick \& Schade (1990) presents cumulative quasar surface densities that have been determined using input from many authors. Down to $B=22.5$ they estimate a population of $130\pm30$ QSO/$\sq\degr$ at $z<2.2$ and $30^{\,+24}_{-12}$ QSO/$\sq\degr$ at $2.2<z<3.3$. Due to the supposed strong decline of the density at $z>3.3$, hardly any contribution was expected from that range. The authors conclude, the surface density for high-redshift quasars were generally lower by a factor of 4 than that of low-redshift objects at similar apparent magnitude.

Hawkins \& Veron (1995) studied a variability selected sample and derived surface densities down to $B=21$, finding a value of \apr 5 QSO/$\sq\degr$ to this limit at $z>2.2$. They suggest to model surface densities as featureless power-law functions of $B$ magnitude. Extrapolating their low-redshift results to fainter $B$ magnitudes is difficult unless Seyfert galaxies are included. From our $z \approx 2.2$-objects, we estimate the faintest $z=2.2$-quasars ($M_B=-23^m$) to have $B \approx 22\fm7$. Therefore, counts of true quasars at $z<2.2$ must saturate towards this magnitude level. For the quasars at $2.2<z<3.2$, we get an estimate of \apr 20 QSO/$\sq\degr$ by extrapolating the power-law model of Hawkins \& Veron down to $B=22.5$.

Hall et al. used data from the Deep Multicolor Survey (DMS) and took spectra of UVX selected quasar candidates down to $B<22.6$, as well as of BRX and VRX selected objects with $V<22$ and $R<22$. This survey covered 0.83 $\sq\degr$ of sky, about 30 CADIS fields. The DMS authors took spectra on a subsample of their candidates. Extrapolation of their confirmation rate to the full number of identified candidates was reported to result in \apr $105\pm30$ QSO/$\sq\degr$ at low redshift and $15\pm4$ QSO/$\sq\degr$ at high redshift, $z>2.3$ \cite{DMS1,DMS2,DMS3}.

At low redshift, $z<2.2$, CADIS can not be expected to shed new light onto quasar counts, since UVX surveys should already find quasars at a high rate of completeness. We do not know how much an additional $U$ band in the CADIS filterset would increase our completeness, but from our Monte-Carlo simulations, we expect to be reasonably complete already with the present filterset. Altogether, our low-redshift results are very consistent with the values reported in the literature.

On the other hand, our first results regarding the density of high-redshift quasars appear to point at values significantly higher than those reported from previous observations (see Tab.\,\ref{density}). They seem to show up with five to ten times higher abundance than in surveys carried out on any other field of the sky, where the observed surface density has been reported to be below one quasar per CADIS field area.

\section{Comparison with luminosity functions}

We do not know perfectly, yet, how incomplete CADIS might be. According to our Monte-Carlo simulations we should be roughly 90\% complete at $R<22^m$. On the other hand, the six-color DMS, e.\,g., found a density of only 15 QSO/$\sq\degr$ at z$>$2.3 with $B<22.6^m$ and $R<22^m$. The DMS authors compared their observations with predictions based on a luminosity function (double power-law model and apparent falloff in space density above $z=3.3$) derived from a quasar survey limited to objects of $R<20^m$ \cite{Warren}. Extrapolating this luminosity function down to the fainter luminosities reached in the DMS predicts about 60 QSO/$\sq\degr$ for $z>2.3$ \cite{Osmer}, implying an incompleteness of about 75\% for the DMS, if the extrapolation is valid. Our own calculation using their luminosity function predicts 68 QSO/$\sq\degr$ at $z>2.2$, the limit most often used for high-redshift quasars.

If applied to the present CADIS data, color excess rules similar to those used in the DMS would detect only one out of the nine z$>$2.2-quasars found in CADIS. Therefore, we just assume a completeness of 100\% for a first comparison with the same luminosity function. The density found in CADIS amounts to roughly 200 QSO/$\sq\degr$, which is still more than three times the value predicted by the above luminosity function. Even though we assumed a detection completeness of 100\%, a strong discrepancy remains. Our advantage in survey completeness can obviously not account for the excess, and out to $z=3.7$, we cannot find any hint for a dramatic decline in the QSO number density. We have carried out a $V/V_{max}$ test \cite{VV} for the 17 objects with $R<22^m$. Using $R=22^m$ as the magnitude that limits the volume for this test, we get in fact
\begin{eqnarray*}
\langle V/V_{max} \rangle = 0.51\pm0.06 ~, 
\end{eqnarray*}
underlining the redshift-independence of the density distribution in our preliminary quasar list. In comparison, a perfectly flat distribution of 17 $V/V_{max}$ values would result in
\begin{eqnarray*}
\langle V/V_{max} \rangle = \frac{1}{2} \pm \sqrt{\frac{1}{12 n}}  = 0.500\pm0.070 ~. 
\end{eqnarray*}

\section{Discussion}

The observations presented here and their comparison with previous work leave us with only two possible alternative interpretations:
\begin{enumerate}

\item Either the abundance of high-redshift quasars has already been estimated quite accurately, and, according to the luminosity function used here, a complete search method should find roughly 70 QSO/$\sq\degr$ at $R<22$ and $z>2.2$. Then, a typical CADIS field should contain on average 1.9 quasars within these limits. In this case, the CADIS 16\,h-field is unusually rich in high-redshift quasars with six objects. Based on Poisson statistics only, it can be shown that only 1.3\% of all CADIS-sized fields should contain more than five quasars of this kind.

\item Or the presented results are a first indication that the luminosity function by Warren et al. (1994) underestimates the abundance of high-redshift quasars significantly, while the counts of quasars at low redshift would remain unaffected by our studies. As a result, the redshift dependence of the luminosity function would be much less pronounced then previously thought. At this point, it is important to keep in mind that we are reporting about not very luminous quasars here ($-27^m < M_B < -23^m$), and results about high-$z$ quasar evolution could depend on the luminosity range investigated.

\end{enumerate}

In any case, to our knowledge the CADIS 16\,h-field is up to date the richest observed field of 100\,$\sq\arcmin$ size in quasars of $R<22^m$ and $z>2.2$. But how could we decide between those two possibilities?

\begin{figure}
\centerline{\hbox{
\psfig{figure=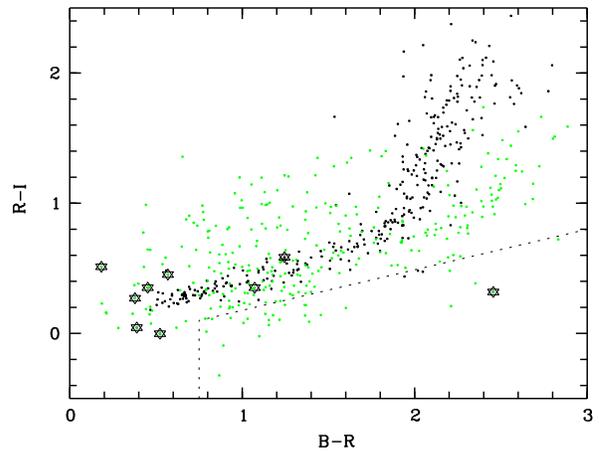,angle=270,clip=t,width=7.8cm}}}
\caption[ ]{This BRI color diagram shows all objects with stellar morphology (unresolved sources) and $R<23^m$ in the 9\,h- and 16\,h-fields. Objects classified as galaxies by the multicolor algorithm are displayed as grey dots, those classified as stars appear as black dots, and the first nine high-redshift quasars ($z>2.2$) identified in CADIS are indicated by stars. Many quasars have stellar broadband colors and could not have been found in multicolor surveys working without mediumband filters. The dashed line corresponds to the BRX selection criterion of the DMS. It would identify only the quasar 16\,h-780 with $z=3.72$ among a number of candidates which are mostly compact galaxies. \label{qso_bri}}
\end{figure}

Color-color diagrams have been useful for object classification before. But when applied to the identification of quasar candidates, the typical color excess rules produce a large number of candidates with strong contamination. Color selection schemes similar to those used in the DMS would recover only a small fraction of the quasars found by CADIS. Most objects found by us would have slipped into the stellar locus in previous multicolor surveys (see Fig.\,\ref{qso_bri}). In the color space of CADIS, half of the quasars became already conspicuous by displaying an emission line in a mediumband filter with more than 0.2 mag excess. The methodogical advantages of CADIS due to its candidate selection scheme support the assumption that many high-redshift quasars have escaped discovery in previous studies, while being detectable for CADIS. 

On the other hand, if we were not able to confirm a comparable number of high-redshift quasars in the other CADIS fields, but instead detected just the complete number of quasars predicted by the luminosity function, we would know, that the 16\,h-field is among the richest fields of CADIS-typical size in the sky. The quickest way to clarify the issue is to complete the follow-up spectroscopy on the 9\,h-field and one more CADIS field where comparable amounts of photometric data are available for the classification.

Whatever the reason is for finding many high-redshift objects, such dense quasar fields offer unique chances for studies of large-scale structure, taking advantage of the fact, that typical lateral distances of the lines of sight to the quasars range between 0.5 and 5 Mpc. Another application is a study of the inverse effect, a powerful tool to estimate the intergalactic radiation field at high redshift, which still needs some observational verification \cite{R2,CF}. The presented results establish that CADIS finds high-redshift quasars with unprecedented completeness. 

\begin{acknowledgements}
The authors thank H.-M. Adorf for helpful discussion about classification methods and their fine-tuning. We also thank the Calar Alto staff for their help and support during many observing runs at the observatory. We thank D. Calzetti for kindly making available the galaxy templates in digital form. The Second Palomar Observatory Sky Survey (POSS-II) was made by the California Institute of Technology with funds from the National Science Foundation Geographic Society, the Sloan Foundation, the Samuel Oschin Foundation, and the Eastman Kodak Corporation.
\end{acknowledgements}

\end{document}